\documentclass[12pt]{article}
\usepackage{amsmath,amssymb,bm,graphicx}
\usepackage{color} 
\usepackage{hyperref}
\usepackage{cite}
\newcommand{\delslash}{\not \! \partial}

\begin{document}

\begin{flushright}
\end{flushright}

\vskip 0.5 truecm

\begin{center}
{\Large{\bf Parity of the neutron  consistent with  neutron-antineutron oscillations 
}}
\end{center}
\vskip .5 truecm
\begin{center}
{\bf { Kazuo Fujikawa$^{1}$ and Anca Tureanu$^2$}}
\end{center}

\begin{center}
\vspace*{0.4cm} 
{\it {$^1$Interdisciplinary Theoretical and Mathematical Sciences Program,\\
RIKEN, Wako 351-0198, Japan\\
\vspace*{0.2cm} 
$^2$Department of Physics, University of Helsinki, P.O.Box 64, 
\\FIN-00014 Helsinki,
Finland
}}
\end{center}
\makeatletter
%\@addtoreset{equation}{section}
%\def\theequation{\thesection.\arabic{equation}}
\makeatother

%\vskip 0.5 truecm

\begin{abstract} 
In the analysis of neutron-antineutron oscillations, it has been recently argued in the literature that the use of the $i\gamma^{0}$ parity $n^{p}(t,-\vec{x})=i\gamma^{0}n(t,-\vec{x})$ which is consistent with the Majorana condition is mandatory and that the ordinary parity transformation of the neutron field $n^{p}(t,-\vec{x}) = \gamma^{0}n(t,-\vec{x})$ has a difficulty. 
We show that a careful treatment of the ordinary parity transformation of the neutron works in the analysis of neutron-antineutron oscillations. Technically, the CP symmetry in the mass diagonalization procedure is important and the two parity transformations, $i\gamma^{0}$-parity and $\gamma^{0}$-parity, are compensated by the Pauli--G\"ursey transformation. Our 
analysis shows that either choice of the parity gives the correct results of neutron-antineutron oscillations if carefully treated. 
\end{abstract}

%\maketitle
%\large

\section{Introduction}

 Motivated by the possible baryon number violation in some unification schemes, the neutron-antineutron oscillations have been discussed by many authors in the past \cite{Kuzmin, Mohapatra,Glashow,  Chang1, Marshak, Kuo, Chang2, Kazarnovsky,Rao-Shrock} (see also the reviews \cite{Mohapatra2, Phillips}) and, in spite of the phenomenon not having been yet observed, experimental bounds have been established \cite{exp1}. The experimental activities are planned to continue in the close future \cite{exp2}. The interest in the theoretical aspects of the discrete symmetries in the context of neutron-antineutron oscillations was recently aroused by the paper \cite{bv}, which was then followed by several related works~\cite{ft, nelson, gardner, FT1, FT2,AT, BV}.  

Historically, it appears that people did not pay much attention to the existence of different definitions of the parity operation or simply used the conventional  $\gamma^{0}$-parity in the analysis of neutron oscillations. In~\cite{FT1}, for example, two definitions of parity were used in the analysis of two different aspects of neutron oscillations and the $i\gamma^{0}$-parity was used in \cite{FT2}.  Recently, Berezhiani and Vainshtein \cite{BV} have performed a detailed analysis of neutron-antineutron oscillations using a two-component spinor notation and the $i\gamma^{0}$-parity. They have shown the consistency of the use of $i\gamma^{0}$-parity in the analysis of neutron-antineutron oscillations. They also commented that the ordinary 
$\gamma^{0}$-parity has a difficulty in the analysis of neutron oscillations. In the present paper, however,  we are going to show that the $\gamma^{0}$-parity for the initial neutron field is perfectly consistent if the neutron oscillations are properly formulated using the CP symmetry for the characterization of the emergent Majorana fermions. Our analysis justifies the common use of the $\gamma^{0}$-parity in neutron-antineutron oscillations in the past. Combined with the analysis of the $i\gamma^{0}$-parity in \cite{BV}, one can thus use either definition of parity in the analysis of neutron oscillations. We emphasize that the known physics related to the neutron, namely, the hadron scattering and the entire nuclear physics, is based on the use of $\gamma^{0}$-parity (i.e. intrinsic parities of neutron and antineutron $\pm1$), and thus the consistent description of neutron-antineutron oscillations by 
$\gamma^{0}$-parity is in fact gratifying.
 
To fix the ideas and conventions, we start from the quadratic effective hermitian Lagrangian for the neutron field $n(x)$ with general small $\Delta B=2$ terms added:
\begin{eqnarray}\label{Lagrangian1}
{\cal L}&=&\overline{n}(x)i\gamma^{\mu}\partial_{\mu}n(x) - m_{D}\overline{n}(x)n(x)\nonumber\\
&-&\frac{1}{2}[m n^{T}(x)Cn(x) + m^{\dagger}\overline{n}(x)C\overline{n}^{T}(x)]\nonumber\\
&-&\frac{1}{2}[m_{5} n^{T}(x)C\gamma_{5}n(x) - m_{5}^{\dagger} \overline{n}(x)C\gamma_{5}\overline{n}^{T}(x)],
\end{eqnarray}
where $m_{D}$ is chosen to be a real positive parameter and $m$ and $m_{5}$ are complex parameters,  very small in absolute value, which break the baryon number symmetry. Our notational conventions follow \cite{Bjorken}, in particular, the charge conjugation matrix is defined by  $C=i\gamma^{2}\gamma^{0}$. The parity-violating and fermion number preserving term $\overline{n(x)}(i\gamma_{5}\delta m) n(x)$ is eliminated by a chiral transformation within the framework of \eqref{Lagrangian1}. We analyze the full implications of \eqref{Lagrangian1} without any further constraints except for the assumption that the magnitudes of $|m|$ and $|m_{5}|$ are very small compared to the neutron mass $m_{D}$.
 It is known \cite{Mohapatra2, Phillips} that the  main aspects of the possible  neutron-antineutron oscillations are described  by the above Lagrangian. 

We define the basic discrete transformation operations based on the free neutron which is assumed to be a Dirac field:
\begin{eqnarray}\label{Dirac}
\int d^{4}x {\cal L}_{D}&=&\int d^{4}x \{\overline{n}(x)i\gamma^{\mu}\partial_{\mu}n(x) - m_{D}\overline{n}(x)n(x)\}.
\end{eqnarray}
 We define the charge conjugation C, which is given by the representation theory of the Clifford algebra, by
\begin{eqnarray}\label{C}
n(x) \rightarrow n^{c}(x)=C\overline{n(x)}^{T}, \ \ \ n^{c}(x) \rightarrow n(x),
\end{eqnarray}
and the parity P defined as the mirror symmetry for a Dirac fermion by the customarily used "$\gamma^{0}$-parity"
\begin{eqnarray}\label{gamma-0-parity}
n(t,\vec{x}) \rightarrow n^{p}(t,-\vec{x}) = \gamma^{0}n(t,-\vec{x}).
\end{eqnarray}
Both C and P thus defined preserve the Dirac Lagrangian \eqref{Dirac} invariant.
The CP transformation rules are defined by 
\begin{eqnarray}\label{CP1}
n(t,\vec{x}) \rightarrow {\cal P}{\cal C}n(t,\vec{x}){\cal C}^{\dagger}{\cal P}^{\dagger}&=&n^{cp}(t,-\vec{x})= (n^{c})^{p}(t,-\vec{x}) = -\gamma^{0}n^{c}(t,-\vec{x}),  \nonumber \\ 
n(t,\vec{x}) \rightarrow {\cal C} {\cal P}n(t,\vec{x}){\cal P}^{\dagger}{\cal C}^{\dagger}&=&n^{pc}(t,-\vec{x})= (n^{p})^{c}(t,-\vec{x}) = \gamma^{0}n^{c}(t,-\vec{x})
\end{eqnarray}
where we used $(n^{c})^{p}(t,-\vec{x}) = C\overline{\gamma^{0}n(t,-\vec{x})}^{T}$.
Thus the ordering is important, but when operated on fermionic fields in a general Lagrangian, which is quadratic in fermions, the ordering is not important. The parity transformation of the charge conjugated fields is:
\begin{eqnarray}
n^{c}(x) \rightarrow \ \ \ (n^{c})^{p}(t,-\vec{x}) = -\gamma^{0}n^{c}(t,-\vec{x}).
\end{eqnarray} 
 This definition of parity amounts to assigning an intrinsic parity $+1$ to the neutron and $-1$ to the antineutron.
 
On the other hand, in the original work of Majorana \cite{Majorana}, the free Majorana fermion was defined by the same action as the Dirac fermion in \eqref{Dirac}
but with {\em purely imaginary} Dirac gamma matrices $\gamma^{\mu}$ . Then the free Dirac equation
\begin{eqnarray}
[i\gamma^{\mu}\partial_{\mu}-m]\psi(x) = 0
\end{eqnarray}
is a real differential equation, and one can impose the reality condition on the solution
\begin{eqnarray}\label{reality condition}
\psi(x)^{\star} = \psi(x),
\end{eqnarray} 
which implies the self-conjugation property under the charge conjugation~\footnote{The pure imaginary condition $\psi^{\star}(x)=-\psi(x)$ is also an allowed solution, but we take \eqref{reality condition} as the primary definition in the present paper. 
}. The conventional parity transformation
$\psi(x) \rightarrow \psi^{p}(t, -\vec{x}) = \gamma^{0}\psi(t, -\vec{x})$ cannot maintain the reality condition \eqref{reality condition} for the purely imaginary $\gamma^{0}$.  Thus the ``$i\gamma^{0}$-parity'' 
\begin{eqnarray}\label{modified parity}
\psi(x) \rightarrow  \psi^{p}(t,-\vec{x}) = i\gamma^{0}\psi(t,-\vec{x})
\end{eqnarray}
is chosen as a natural parity transformation rule for the Majorana fields \cite{Majorana, Kayser1}.  

In the generic representation of the Dirac matrices \cite{Bjorken}, the ``$i\gamma^{0}$-parity''  satisfies the condition 
\begin{eqnarray}\label{consistency condition}
i\gamma^{0}\psi(t,-\vec{x}) = C\overline{i\gamma^{0}\psi(t,-\vec{x})}^{T}
\end{eqnarray}
for the field which satisfies the classical Majorana condition 
\begin{eqnarray}\label{classical Majorana condition}
\psi (x) = C\overline{\psi(x)}^{T}
\end{eqnarray}
and thus  $i\gamma^{0}$-parity is a natural choice of the parity for the Majorana fermion also in this generic representation  (as well as in any other). See \cite{Weinberg2} for the phase freedom in the definition of the parity operation.
The $i\gamma^{0}$-parity can be used for free Dirac fields as well, with the mention that the intrinsic parity assigned to the corresponding particle and antiparticle is the same, $i$.

For the Dirac fermion with $U(1)$ fermion number freedom, these two definitions of parity are equivalent, but their equivalence is not obvious for theories with broken fermionic number such as \eqref{Lagrangian1}. One may rather 
suspect that the conventional $\gamma^{0}$-parity is inconsistent in theories where Majorana fermions appear. 
In the analysis of neutron-antineutron oscillations described by \eqref{Lagrangian1}, one visualizes a virtual process where the initial neutron turns into a superposition of non-degenerate Majorana fermions which after oscillations ends up to be an antineutron. If one uses the $\gamma^{0}$-parity operation for the starting neutron, one may thus suspect that a contradiction appears for the intermediate states with Majorana fermions.
On the other hand,  the $i\gamma^{0}$-parity is consistent with both Dirac and Majorana fermions and thus intuitively more natural for  neutron-antineutron oscillations.
% and, in fact, it is known to be the case as explicitly demonstrated in \cite{BV}.

In the present paper, however, we are going to show that the use of the conventional $\gamma^{0}$-parity for the starting neutron field gives a logically consistent description of neutron-antineutron oscillations if a proper treatment and interpretation is applied. The basic idea leading to this statement is that C and P defined for the free Dirac Lagrangian \eqref{Dirac} as described above are not generally well defined after the mass diagonalization of the general Lagrangian \eqref{Lagrangian1} (see, however, the Appendix \ref{A} later), but the CP symmetry is defined for the general Lagrangian \eqref{Lagrangian1} after the diagonalization of the mass terms; the mass diagonalization is after all required to define Majorana fermions. The proposal in~\cite{Fujikawa} is then to characterize the emergent Majorana fermions by the CP symmetry~\footnote{ It has been recently shown~\cite{Fujikawa1} that the use of either $\gamma^{0}$-parity or $i\gamma^{0}$-parity for the chiral fermions gives the consistent equivalent description of emergent Majorana fermions in Weinberg's model of neutrinos~\cite{Weinberg} in an extension of the Standard Model, when CP is used to characterize the Majorana neutrino formed of chiral fermions. }.  A formal proof of the canonical equivalence
of the two choices of the parity operation in the analysis of neutron-antineutron oscillations shall be given in the Appendix \ref{A} using the Pauli--G\"ursey transformation.

\section{Neutron-antineutron oscillations  with $\gamma^{0}$-parity}

\subsection{Consistent description of Majorana fermions}
We first rewrite the hermitian Lagrangian \eqref{Lagrangian1} in terms of chiral notations as
\begin{eqnarray}\label{Lagrangian-2}
{\cal L}&=&\overline{n_{L}}(x)i\gamma^{\mu}\partial_{\mu}n_{L}(x) +
\overline{n_{R}}(x)i\gamma^{\mu}\partial_{\mu}n_{R}(x)\nonumber\\
&-& m_{D}\overline{n_{L}}(x)n_{R}(x) 
-\frac{1}{2}m_{L}n_{L}^{T}(x)Cn_{L}(x) 
-\frac{1}{2}m_{R}n_{R}^{T}(x)Cn_{R}(x) + h.c.,
\end{eqnarray}
with $n_{R,L}(x)=[(1\pm \gamma_{5})/2]n(x)$, for the effective use of CP transformation to characterize the Majorana fermions. In terms of the mass parameters in \eqref{Lagrangian1}
\begin{eqnarray}\label{mass parameters1}
m_{D}, \ \ \ m_{L}= m-m_{5},\ \ \ m_{R}= m+m_{5},
\end{eqnarray} 
namely, we define complex mass parameters $m_{L}$ and $m_{R}$, while $m_{D}$ is chosen to be real. 

  We recall the transformation laws of chiral fermions derived from the chiral projection of the Dirac fermionic field:
\begin{eqnarray}\label{conventional C and P}
&&C:\ n_{L}(x) \rightarrow C\overline{n_{R}(x)}^{T},\ \ \ n_{R}(x) \rightarrow C\overline{n_{L}(x)}^{T},\nonumber\\
&&P:\ n_{L}(x) \rightarrow  \gamma^{0}n_{R}(t,-\vec{x}),\ \ \ n_{R}(x) \rightarrow  \gamma^{0}n_{L}(t,-\vec{x}),\nonumber\\
&&CP:\ n_{L}(x) \rightarrow -\gamma^{0}C\overline{n_{L}(t,-\vec{x})}^{T},\ \ \ n_{R}(x) \rightarrow -\gamma^{0}C\overline{n_{R}(t,-\vec{x})}^{T}.
\end{eqnarray}  
The minus signature in CP transformation shows that we use the $n^{cp}(t,-\vec{x})$ convention in \eqref{CP1}.  The baryon number violating mass terms in the Lagarangian \eqref{Lagrangian-2} are transformed under C, P and CP transformations in \eqref{conventional C and P}
as 
\begin{eqnarray}
C&:&
-\frac{1}{2}m_{L}n_{L}^{T}(x)Cn_{L}(x) 
-\frac{1}{2}m_{R}n_{R}^{T}(x)Cn_{R}(x) + h.c.
\nonumber\\
&\rightarrow& 
-\frac{1}{2}m^{\dagger}_{R}n_{L}^{T}(x)Cn_{L}(x)
-\frac{1}{2}m^{\dagger}_{L}n_{R}^{T}(x)Cn_{R}(x) + h.c.
\nonumber\\
P&:& 
-\frac{1}{2}m_{L}n_{L}^{T}(x)Cn_{L}(x) 
-\frac{1}{2}m_{R}n_{R}^{T}(x)Cn_{R}(x) + h.c.
\nonumber\\
&\rightarrow&
+\frac{1}{2}m_{L}n_{R}^{T}(x)Cn_{R}(x) 
+\frac{1}{2}m_{R}n_{L}^{T}(x)Cn_{L}(x) + h.c.
\nonumber\\
CP&:&
-\frac{1}{2}m_{L}n_{L}^{T}(x)Cn_{L}(x) 
-\frac{1}{2}m_{R}n_{R}^{T}(x)Cn_{R}(x) + h.c.
\nonumber\\
&\rightarrow&
+\frac{1}{2}m^{\dagger}_{L}n_{L}^{T}(x)Cn_{L}(x) 
+\frac{1}{2}m^{\dagger}_{R}n_{R}^{T}(x)Cn_{R}(x) + h.c.
\end{eqnarray}
Namely, in the Lagrangian \eqref{Lagrangian-2}, C is a good symmetry for $m_{L}=m^{\dagger}_{R}$ and P is a good symmetry for $m_{L}=-m_{R}$; CP is a good symmetry for
\begin{eqnarray}
m_{L} = -m_{L}^{\dagger}  \ \ {\rm  and } \ \  m_{R}= -m_{R}^{\dagger},
\end{eqnarray}
which can hold without any relation between $m_{L}$ and $m_{R}$. We however emphasize that these symmetry properties of the ``bare parameters'' have no definite meanings after the mass diagonalization in general, since the canonical transformation parameterized by the matrix $U$ which diagonalizes the mass terms to define Majorana fermions carries away these discrete symmetries from the sector of the fermion mass terms to the interaction terms, as in the Standard Model. The actual discrete symmetries are better analyzed based on the Lagrangian after the mass diagonalization. On the other hand, one may want to know the more detailed direct physical implications of the starting Lagrangian based on the direct diagonalization. This aspect is briefly mentioned in the Appendix \ref{B}.

We next diagonalize the Lagrangian \eqref{Lagrangian-2} by writing the mass term as 
\begin{eqnarray}
(-2){\cal L}_{mass}=
\left(\begin{array}{cc}
            \overline{n_{R}}&\overline{n_{R}^{c}}
            \end{array}\right)
\left(\begin{array}{cc}
            m^{\dagger}_{R}& m_{D}\\
            m_{D}&m_{L}
            \end{array}\right)
            \left(\begin{array}{c}
            n_{L}^{c}\\
            n_{L}
            \end{array}\right) +h.c.,
\end{eqnarray}
where we defined 
\begin{eqnarray}
n_{L}^{c}\equiv C\overline{n_{R}}^T, \ \ \ n_{R}^{c}\equiv C\overline{n_{L}}^T . 
\end{eqnarray}
We diagonalize the {\em complex symmetric} mass matrix 
using a $2 \times 2$ unitary matrix (Autonne--Takagi factorization)  \cite{Autonne, Takagi}          
\begin{eqnarray}\label{diagonal mass-1}
            U^{T}
            \left(\begin{array}{cc}
            m^{\dagger}_{R}& m_{D}\\
            m_{D}& m_{L}
            \end{array}\right)
            U
            = i \left(\begin{array}{cc}
            M_{1}&0\\
            0&M_{2}
            \end{array}\right)    ,        
\end{eqnarray}
where  $M_{1}$ and $M_{2}$ are  positive real numbers, which can be chosen to be the characteristic values~\footnote{The relation \eqref{diagonal mass-1} may be regarded as the ordinary Autonne--Takagi factorization of the matrix $(-i)\left(\begin{array}{cc}
            m^{\dagger}_{R}& m_{D}\\
            m_{D}& m_{L}
            \end{array}\right)$. Also, the relation \eqref{diagonal mass-1} written in terms of a  unitary matrix  $U e^{-i\pi/4}$ shall be  discussed in Appendix \ref{A} in connection with the Pauli--G\"ursey transformation. Mathematically, the Autonne--Takagi factorization with a suitable unitary matrix gives rise to characterisitic values, and thus $M_{1}$ and $M_{2}$ can be chosen to be real and positive for a suitable $U e^{-i\pi/4}$ in \eqref{diagonal mass-1}. A crucial difference between the Autonne--Takagi factorization and the conventional unitary transformation is that one can change the phase of the diagonal elements freely by choosing a suitable $U$ such as in \eqref{diagonal mass-1}. }.
This form of diagonalization makes the Lagrangian with diagonal mass matrix CP invariant.

When one defines
\begin{eqnarray} \label{variable-change-1}          
            &&\left(\begin{array}{c}
            n_{L}^{c}\\
            n_{L}
            \end{array}\right)
            = U \left(\begin{array}{c}
            {N}_{L}^{c}\\
            {N}_{L}
            \end{array}\right)
           ,\ \ \ \ 
            \left(\begin{array}{c}
            n_{R}\\
            n_{R}^{c}
            \end{array}\right)
            = U^{\star} 
            \left(\begin{array}{c}
            {N}_{R}\\
            {N}_{R}^{c}
            \end{array}\right)          
\end{eqnarray}
the mass term of the Lagrangian \eqref{Lagrangian-2} becomes
\begin{eqnarray}\label{exact-mass0}
(-2){\cal L}_{mass}
&=& i\left(\begin{array}{cc}
            \overline{{N}_{R}}&\overline{{N}_{R}^{c}}
            \end{array}\right)
\left(\begin{array}{cc}
            M_{1}&0\\
            0&M_{2} 
            \end{array}\right)            
            \left(\begin{array}{c}
            {N}_{L}^{c}\\
            {N}_{L}
            \end{array}\right) +h.c.                 
\end{eqnarray}
The total hermitian Lagrangian is then written as  
\begin{eqnarray}\label{exact-solution0}
{\cal L}&=&\frac{1}{2}\{\overline{N_{L}}(x)i\delslash N_{L}(x)+\overline{N^{c}_{L}}(x)i\delslash N^{c}_{L}(x)+\overline{N_{R}}(x)i\delslash N_{R}(x)
+\overline{N^{c}_{R}}(x)i\delslash N^{c}_{R}(x)\}\nonumber\\
&-&(i/2)\left(\begin{array}{cc}
            \overline{N_{R}}&\overline{N_{R}^{c}}
            \end{array}\right)
\left(\begin{array}{cc}
            M_{1}&0\\
            0&M_{2} 
            \end{array}\right)            
            \left(\begin{array}{c}
            N_{L}^{c}\\
            N_{L}
            \end{array}\right) +h.c.,       \nonumber\\
&=&\overline{N_{L}}(x)i\delslash N_{L}(x)+\overline{N_{R}}(x)i\delslash N_{R}(x)\nonumber\\
&-&(i/2)\{\overline{N_{R}}CM_{1}\overline{n_{R}}^{T}-N_{R}^{T}CM_{1}N_{R} -\overline{N_{L}}CM_{2}\overline{N_{L}}^{T}+N_{L}^{T}CM_{2}N_{L}
\} \nonumber\\
&=&\frac{1}{2} \overline{\psi_{+}(x)}(i\gamma^{\mu}\partial_{\mu} - M_{1})\psi_{+}(x) + 
\frac{1}{2}\overline{\psi_{-}(x)}(i\gamma^{\mu}\partial_{\mu} - M_{2})\psi_{-}(x),
\end{eqnarray}
where we defined the Majorana fields by 
\begin{eqnarray}\label{Majorana}
\psi_{+}(x)&=& e^{i\pi/4}N_{R}(x) - e^{-i\pi/4}C\overline{N_{R}(x)}^{T}, \nonumber\\
\psi_{-}(x)&=& e^{i\pi/4}N_{L}(x) + e^{-i\pi/4}C\overline{N_{L}(x)}^{T} ,
\end{eqnarray}
which satisfy the classical Majorana conditions
\begin{eqnarray}\label{classical Majorana}
\psi_{+}(x)= -C\overline{\psi_{+}(x)}^{T}, \ \ \ \psi_{-}(x)=  C\overline{\psi_{-}(x)}^{T},
\end{eqnarray}
identically in the sense that these conditions are satisfied regardless of the choice of $N_{R}(x)$ or $N_{L}(x)$; one may replace $N_{R}(x)$ by an arbitrary fermion field $f_{R}(x)$ in \eqref{Majorana}, for example, and the resulting $\psi_{+}(x)$ still satisfies the condition \eqref{classical Majorana}. We take the relations \eqref{classical Majorana} combined with the Dirac equations 
\begin{eqnarray}
(i\gamma^{\mu}\partial_{\mu} - M_{1})\psi_{+}(x)=0, \ \ \ (i\gamma^{\mu}\partial_{\mu} - M_{2})\psi_{-}(x)=0
\end{eqnarray}
as the definition of Majorana fermions based on the analysis of the Clifford algebra. One may try to define the Majorana fermion defined in \eqref{Majorana} using a non-trivial charge conjugation operator, but such an attempt generally fails \cite{FT-1}. See, however, Subsection \ref{2.3} later and the Appendix \ref{A}. 

The transformation \eqref{variable-change-1}  preserves the form of the kinetic term in the Lagrangian and thus the canonical anti-commutators; the transformation thus constitutes a {\em canonical transformation}. The discrete symmetry rules \eqref{conventional C and P} are thus applied to new variables every time after the canonical transformation~\cite{Pauli, Gursey, KF-PG} \footnote{We apply the same transformation rules to the old variables also. Thus the C,  P and CP transformations of the old variables do not reproduce in general those symmetries of the new variables after the canonical transformation, which changes the forms of the mass terms and interaction terms.  The classic Kobayashi-Maskawa analysis of CP violation illustrates an example of the use of a canonical transformation \cite{KM}.}.  As will be  explained in the Appendix \ref{A}, the $U(2)$ transformation \eqref{variable-change-1} is related to the Pauli--G\"ursey transformation.
 
 The Lagrangian \eqref{exact-solution0} is not invariant under the C nor P transformation defined conventionally in \eqref{conventional C and P} separately for $M_{1}\neq M_{2}$ as is shown below, while the Lagrangian is invariant under the CP transformation in \eqref{conventional C and P}, 
$N_{L}(x) \rightarrow -\gamma^{0}C\overline{N_{L}(t,-\vec{x})}^{T}$ and $N_{R}(x) \rightarrow -\gamma^{0}C\overline{N_{R}(t,-\vec{x})}^{T}$,
 for any $M_{1}$ and $M_{2}$. 
We can also confirm  using the formal operator notations for the transformations  \eqref{conventional C and P}
\begin{eqnarray}\label{CP of emergent Majorana}
&&({\cal P}{\cal C})\{ e^{i\pi/4}N_{R}(x) - e^{-i\pi/4}C\overline{N_{R}(x)}^{T} \}({\cal P}{\cal C})^{\dagger} = i\gamma^{0}\psi_{+}(t,-\vec{x})     \nonumber\\
&&({\cal P}{\cal C})\{ e^{i\pi/4}N_{L}(x) + e^{-i\pi/4}C\overline{N_{L}(x)}^{T} \}({\cal P}{\cal C})^{\dagger} = -i\gamma^{0}\psi_{-}(t,-\vec{x}).
\end{eqnarray}
We thus characterized  these Majorana fields by CP symmetry in \eqref{conventional C and P}
\begin{eqnarray}\label{CP of Majorana-1}
({\cal P}{\cal C})\psi_{+}(x)({\cal P}{\cal C})^{\dagger} &=& -i\gamma^{0} C\overline{\psi_{+}(t,-\vec{x})}^{T} = i\gamma^{0}\psi_{+}(t,-\vec{x}), \nonumber\\
({\cal P}{\cal C})\psi_{-}(x)({\cal P}{\cal C})^{\dagger} &=&- i\gamma^{0} C\overline{\psi_{-}(t,-\vec{x})}^{T} = - i\gamma^{0}\psi_{-}(t,-\vec{x}),
\end{eqnarray} 
where the first equalities are the operator relations while the second equalities are the classical Majorana conditions \eqref{classical Majorana}; these two operations combined reproduce \eqref{CP of emergent Majorana}.  It is crucial that these CP transforms are consistent with the classical Majorana conditions in the sense
\begin{eqnarray}\label{consistency condition}
&&\psi_{+}(x)= -C\overline{\psi_{+}(x)}^{T}\ \rightarrow\ i\gamma^{0}\psi_{+}(t,-\vec{x}) = -C\overline{i\gamma^{0}\psi_{+}(t,-\vec{x})}\nonumber\\
&&\psi_{-}(x)=  C\overline{\psi_{-}(x)}^{T} \ \rightarrow\ -i\gamma^{0}\psi_{-}(t,-\vec{x}) = C\overline{-i\gamma^{0}\psi_{-}(t,-\vec{x})}
\end{eqnarray}
which is a counter part of the crucial consistency of the $i\gamma^{0}$-parity \eqref{modified parity} and the classical Majorana condition \eqref{classical Majorana condition}.  In the present formulation, we do not specify the parity transformation for the Majorana field, but specify the CP-parity. One can see that, while the CP-transformation \eqref{CP of Majorana-1} is specified and leaves the Lagrangian invariant, on the other hand we have for the $\gamma_0$-parity:
\begin{eqnarray}\label{parity of Majorana}
{\cal P}\psi_{+}(x){\cal P}^{\dagger} &=& {\cal P}\{e^{i\pi/4}N_{R}(x) - e^{-i\pi/4}C\overline{N_{R}(x)}^{T}\}{\cal P}^{\dagger} \nonumber\\
&=& e^{i\pi/4}\gamma^{0}N_{L}(t,-\vec{x}) - e^{-i\pi/4}C\overline{\gamma^{0}N_{L}(t,-\vec{x})}^{T}\nonumber\\
&=&\gamma^{0}[ e^{i\pi/4}N_{L}(t,-\vec{x}) + e^{-i\pi/4}C\overline{N_{L}(t,-\vec{x})}^{T}]\nonumber\\
&=& \gamma^{0}\psi_{-}(t,-\vec{x}),\nonumber\\
{\cal P}\psi_{-}(x){\cal P}^{\dagger} &=&\gamma^{0}\psi_{+}(t,-\vec{x}),
\end{eqnarray} 
which is not a symmetry of the Lagrangian \eqref{exact-solution0} for $M_{1}\neq M_{2}$ which is required for neutron oscillations.  

We emphasize that the relations \eqref{parity of Majorana} correspond to the ``{\em parity doubling theorem}'' we discussed before, namely, the $\gamma^{0}$-parity invariance of the Lagrangian after mass diagonalization
leads to the degeneracy of the emergent Majorana fermions $M_{1} = M_{2}$ and thus to no neutron-antineutron oscillations \cite{FT1}. In this sense, the $\gamma^0$-parity is a criterion of discrimination between the Lagrangians which may lead to oscillations and those that do not: $\gamma^0$-parity violation of the baryon number violating Lagrangian written in terms of the original neutron field is a necessary condition for oscillation (see eq.\eqref{Mass eigenvlaues4} later). Note that this is a technical criterion and it does not mean a physically observable parity violation in oscillation, as explained in \cite{FT1}. 

The charge conjugation is not a symmetry  either:
\begin{eqnarray}
 {\cal C}\psi_{+}(x){\cal C}^{\dagger} &=& {\cal C}\{e^{i\pi/4}N_{R}(x) - e^{-i\pi/4}C\overline{N_{R}(x)}^{T}\}{\cal C}^{\dagger} \nonumber\\
&=& e^{i\pi/4}C\overline{N_{L}}^{T}(x) - e^{-i\pi/4}N_{L}(x)\nonumber\\
&=&i[ e^{i\pi/4}N_{L}(x) + e^{-i\pi/4}C\overline{N_{L}(x)}^{T}]\nonumber\\
&=& i\psi_{-}(x),\nonumber\\
{\cal C}\psi_{-}(x){\cal C}^{\dagger} &=& -i\psi_{+}(x),
\end{eqnarray} 
namely, not the symmetry of the Lagrangian \eqref{exact-solution0} for $M_{1}\neq M_{2}$.

 \subsection{Neutron-antineutron oscillations}
 The starting neutron field, which is understood as the neutron produced by strong interactions, is related to the ``mass eigenstate'' ${N}(x)$ by \eqref{variable-change-1}  that is in turn expressed in terms of Majorana fields. 
We define the new fields
\begin{eqnarray}\label{variables with hat}
\hat{n}(x) \equiv  e^{i\pi/4}n(x), \ \ \   \hat{N}(x) \equiv  e^{i\pi/4}N(x),  
\end{eqnarray}
in terms of which the Majorana fields \eqref{Majorana} are written as  
\begin{eqnarray}\label{Majorana2}
\psi_{+}(x)&=& \hat{N}_{R}(x) - C\overline{\hat{N}_{R}(x)}^{T}, \nonumber\\
\psi_{-}(x)&=& \hat{N}_{L}(x) + C\overline{\hat{N}_{L}(x)}^{T} 
\end{eqnarray}
and we have
\begin{eqnarray}\label{original field-1}
\hat{n}(x)&=&\hat{n}_{R}+\hat{n}_{L} \nonumber\\
&=& (\hat{U}^{\star}_{11}\psi_{+}(x)_{R} - \hat{U}_{21} \psi_{+}(x)_{L}) +(\hat{U}^{\star}_{12}\psi_{-}(x)_{R} +\hat{U}_{22} \psi_{-}(x)_{L}), \nonumber\\
\hat{n}^{c}(x)&=&\hat{n}^{c}_{R}+\hat{n}^{c}_{L}\nonumber\\
&=& (\hat{U}^{\star}_{21}\psi_{+}(x)_{R} - \hat{U}_{11} \psi_{+}(x)_{L}) + (\hat{U}^{\star}_{22}\psi_{-}(x)_{R} + \hat{U}_{12} \psi_{-}(x)_{L}).
\end{eqnarray}
We defined  the matrix elements of  a new $2\times 2$ unitary matrix 
\begin{eqnarray}\label{matrix with hat}
\hat{U}&\equiv&\left(\begin{array}{cc}
            e^{-i\pi/4}& 0\\
            0& e^{i\pi/4}
            \end{array}\right)U\left(\begin{array}{cc}
            e^{i\pi/4}& 0\\
            0& e^{-i\pi/4}
            \end{array}\right)\nonumber\\
            &=&\left(\begin{array}{cc}
            e^{-i\pi/4}& 0\\
            0& e^{i\pi/4}
            \end{array}\right)\left(\begin{array}{cc}
            U_{11}& U_{12}\\
            U_{21}& U_{22}
            \end{array}\right)\left(\begin{array}{cc}
            e^{i\pi/4}& 0\\
            0& e^{-i\pi/4}
            \end{array}\right)
\end{eqnarray}
which satisfies instead of \eqref{variable-change-1}
\begin{eqnarray} \label{variable-change-2}          
            &&\left(\begin{array}{c}
            \hat{n}_{L}^{c}\\
            \hat{n}_{L}
            \end{array}\right)
            = \hat{U} \left(\begin{array}{c}
            \hat{N}_{L}^{c}\\
            \hat{N}_{L}
            \end{array}\right)
           ,\ \ \ \ 
            \left(\begin{array}{c}
            \hat{n}_{R}\\
            \hat{n}_{R}^{c}
            \end{array}\right)
            = \hat{U}^{\star} 
            \left(\begin{array}{c}
            \hat{N}_{R}\\
            \hat{N}_{R}^{c}
            \end{array}\right) .         
\end{eqnarray}
The external fields $\hat{n}(x)$ and $\hat{n}^{c}(x)$ are treated as analogues of ``flavor`` fields in the present  neutron-antineutron oscillations~\footnote{If one adjusts the phase conventions of the starting neutron fields $n(x)$ and $n^{c}(x)$ suitably in \eqref{Lagrangian1}, one can avoid the use of fields with the hat-notation. We however prefer to keep the present notational convention to emphasize that we start with a generic Lagrangian \eqref{Lagrangian1} and examine what happens if one applies the conventional $\gamma^{0}$ parity operation.}. When one uses the (valid) CP symmetry of Majorana fermions in \eqref{CP of Majorana-1}, it is confirmed that 
the relations \eqref{original field-1} show that CP is broken for $\hat{U}\neq \hat{U}^{\star}$ in the sense  
\begin{eqnarray}\label{CP invariance}
{\cal CP}\hat{n}(x)({\cal CP})^{\dagger} \neq -i\gamma^{0}\hat{n}^{c}(t, -\vec{x}),
\end{eqnarray}
namely, the operations of CP at the level of Majorana fermions do not agree with the expected operations of CP at the level of the neutron $\hat{n}(x)$ produced by strong interactions (using the definition of \eqref{variables with hat} and the transformations rules \eqref{conventional C and P}).

The unitary matrix $U$ in \eqref{variable-change-1} (or the matrix $\hat{U}$ \eqref{matrix with hat} among the variables with a hat) transfers the CP violating effects to the interaction terms, which contains the coupling to other particles such as the proton depending on the detailed specification of the effective model, leaving the CP invariant Lagrangian \eqref{exact-solution0} for the sector of Majorana fermions.  Unlike the  CP analysis  of the seesaw model in an extension of the Standard Model \cite{ Fukugita, Giunti, Bilenky, FT-1}, which is described by a  Lagrangian closely related to \eqref{Lagrangian-2},  the present effective theory is not designed to analyze the CP symmetry breaking, since the weak current, which describes the transition between $n(x)$ and $p(x)$, is not purely left-handed and $n_{L}$ is a superposition of mass eigenstates $N_{L}$ and $N_{L}^{c}$ with  approximately equal weight factors. This case is very different from SM. A realistic analysis of CP symmetry breaking related to the transition  between $n(x)$ and $p(x)$ would require a study of the fundamental quark level dynamics.
            
On the other hand, the present effective theory is useful to understand a general qualitative aspect of CP symmetry such as the question if CP symmetry can be measured in neutron-antineutron  oscillations by treating $n(x)$ and $n^{c}(x)$ as analogues of flavor fields.  We here discuss this aspect of CP symmetry.

As for the neutron-antineutron oscillations, assuming {\em a sudden projection treatment} (``sudden'' in the sense of the change of the description in terms of a neutron to the description in terms of non-degenerate Majorana fermions), we have  by defining the neutron state at $t=0$ as
$|{n}(0,\vec{p})\rangle={\hat{n}^{\dagger}(0,\vec{p})}^{T}|0\rangle$ and similarly the antineutron state at the time $t$ as  $\langle {\bar n}(t,\vec{p})|=\langle 0|\hat{n}^{c}(t,\vec{p})^{T}$,
\begin{eqnarray}
\langle \bar{n}(t,\vec{p})|{n}(0,\vec{p})\rangle&=&
(\hat{U}^{\star}_{21}\hat{U}_{11})[\langle \psi_{+R}(t,\vec{p})|\psi_{+R}(0,\vec{p})\rangle + \langle \psi_{+L}(t,\vec{p})|\psi_{+L}(0,\vec{p})]\rangle\\
&+&
(\hat{U}^{\star}_{22}\hat{U}_{12})[\langle \psi_{-R}(t,\vec{p})|\psi_{-R}(0,\vec{p})\rangle + \langle \psi_{-L}(t,\vec{p})|\psi_{-L}(0,\vec{p})\rangle]\nonumber\\
&=&(\hat{U}^{\star}_{21}\hat{U}_{11})\langle \psi_{+}(t,\vec{p})|\psi_{+}(0,\vec{p})\rangle + 
(\hat{U}^{\star}_{22}\hat{U}_{12})\langle \psi_{-}(t,\vec{p})|\psi_{-}(0,\vec{p})\rangle.\nonumber
\end{eqnarray}
If one notes the relation
$\hat{U}_{21}\hat{U}^{\star}_{11} + \hat{U}_{22}\hat{U}^{\star}_{12}=0$
arising from the unitarity of $\hat{U}$, one obtains
\begin{eqnarray}\label{oscillation-1}
&&|\langle \bar{n}(t,\vec{p})|{n}(0,\vec{p})\rangle|^{2}\nonumber\\
&&= |(\hat{U}_{21}\hat{U}^{\star}_{11})|^{2}|[\langle \psi_{+}(t,\vec{p})|\psi_{+}(0,\vec{p})\rangle - 
\langle \psi_{-}(t,\vec{p})|\psi_{-}(0,\vec{p})\rangle]|^{2}\nonumber\\
&&=|(\hat{U}_{21}\hat{U}^{\star}_{11})|^{2}|[e^{iE_{1}t}\langle \psi_{+}(0,\vec{p})|\psi_{+}(0,\vec{p})\rangle -  e^{iE_{2}t}
\langle \psi_{-}(0,\vec{p})|\psi_{-}(0,\vec{p})\rangle]|^{2}
\nonumber\\
&&=4|(\hat{U}_{21}\hat{U}^{\star}_{11})|^{2}|\sin^{2} (\Delta E t/2),
\end{eqnarray}
where 
$\Delta E=E_{1}-E_{2}$, with $E_i=\sqrt{\vec p ^2+M_i^2},\  i=1,2$
 and we used $\langle \psi_{+}(0,\vec{p})|\psi_{+}(0,\vec{p})\rangle=\langle \psi_{-}(0,\vec{p})|\psi_{-}(0,\vec{p})\rangle=1$. It is significant that the amplitude $\langle \bar{n}(t,\vec{p})|{n}(0,\vec{p})\rangle$ is expressed in terms of the well-defined $\langle \psi_{+}(t,\vec{p})|\psi_{+}(0,\vec{p})\rangle$ and $\langle \psi_{-}(t,\vec{p})|\psi_{-}(0,\vec{p})\rangle$ without any chiral projection operators  in the present treatment. Hence, the use of the chiral fermions is a matter of mathematical convenience.

The above formula \eqref{oscillation-1} shows that the effect of CP breaking does not appear in the oscillation probability in the present effective theory although the absolute values of the amplitude depend on the possible CP breaking, in agreement with the conclusion in \cite{FT1}. This has been confirmed by a detailed calculation using a quantum field theoretical procedure using the method of unitarily inequivalent representations in Hamiltonian formalism \cite{AT}. 

For the sake of completeness, we here present an exact mass difference after the mass diagonalization \eqref{diagonal mass-1}. We first rewrite \eqref{diagonal mass-1} in the form
\begin{eqnarray}\label{Mass eigenvlaues1}
 U^{\dagger}
            \left(\begin{array}{cc}
            m^{2}_{D}+|m_{R}|^{2}& m_{D}(m_{R}+m_{L})\\
            m_{D}(m^{\dagger}_{R}+m^{\dagger}_{L})        &m_{D}^{2}+ |m_{L}|^{2}
            \end{array}\right)U
            =  \left(\begin{array}{cc}
            M^{2}_{1}&0\\
            0&M^{2}_{2}
            \end{array}\right).    
\end{eqnarray}
From the considerations of the trace and the determinant of this relation, we have 
\begin{eqnarray}\label{Mass eigenvlaues2}
&&M_{1}^{2}+M_{2}^{2}=2m_{D}^{2} + |m_{R}|^{2} + |m_{L}|^{2},\nonumber\\
&&M_{1}^{2}M_{2}^{2}=(m_{D}^{2} + |m_{R}|^{2})(m_{D}^{2} + |m_{L}|^{2})-m_{D}^{2}|m_{R}+m_{L}|^{2}
\end{eqnarray} 
and thus 
\begin{eqnarray}\label{Mass eigenvlaues13}
(M_{1}^{2} - M_{2}^{2})^{2}&=&4m^{2}_{D}|m_{R}+m_{L}|^{2}+(|m_{R}|^{2}-|m_{L}|^{2})^{2}\nonumber\\
&=& 16m_{D}^{2}|m|^{2}+ 4(mm_{5}^{\star}+m^{\star}m_{5})^{2}
\end{eqnarray}
which implies $|M_{1}^{2} - M_{2}^{2}|=4m_{D}|m|$ for $m_{D}\gg |m_{5}|$ for arbitrary $m\neq 0$. Finally 
\begin{eqnarray}\label{Mass eigenvlaues4}
|M_{1} - M_{2}|=2|m|
\end{eqnarray}
for $m_{D}\gg |m|$. In the same approximation, one has $M_{1,2}=m_{D}\pm |m|$ by choosing $M_{1}>M_{2}$. 
%.

The complete absence of CP breaking implies $\hat{U}=\hat{U}^{\star}$ in \eqref{CP invariance}, namely, one may choose a  real unitary $\hat{U}$ in \eqref{matrix with hat} which is a generic orthogonal matrix:
\begin{eqnarray}\label{U-matrix1}
\hat{U}=\left(\begin{array}{cc}
            \cos\theta& \sin\theta\\
            -\sin\theta& \cos\theta
            \end{array}\right).
\end{eqnarray}
Thus in the absence of CP violation contained in  $\hat{U}$, we have the standard formula for the neutron-antineutron oscillation probability  (for nonrelativistic neutrons, with $m_D \gg |m|$ as in \eqref{Mass eigenvlaues4}):
\begin{eqnarray}
|\langle \bar{n}_{0}(t,\vec{p})|{n}_{0}(0,\vec{p})\rangle|^{2}=\sin^{2}(2\theta)\sin^{2} (\Delta E t/2).
\end{eqnarray}

\subsection{Deformed symmetry generated by ${\cal C}_{M}$ and ${\cal P}_{M}$}\label{2.3}
One may wonder if it is possible to define C and P symmetries valid for the emergent Majorana fermions in the present formulation.  It is possible to define a formal {\em deformed symmetry} generated by  \cite{Fujikawa, KF-PG}
\begin{eqnarray} \label{Deformed C and P-2} 
{\cal C}_{M}=1, \ \ \ {\cal P}_{M}={\cal P}{\cal C},
\end{eqnarray}
which is a symmetry of \eqref{exact-solution0} and
\begin{eqnarray}\label{trivial C and CP2}
&&{\cal C}_{M}\psi_{+}(x){\cal C}_{M}^{\dagger}=\psi_{+}(x),\ \ \ {\cal P}_{M}\psi_{+}(x){\cal P}_{M}^{\dagger}= i\gamma^{0}\psi_{+}(t,-\vec{x}),\nonumber\\
&&{\cal C}_{M}\psi_{-}(x){\cal C}_{M}^{\dagger}=\psi_{-}(x),\ \ \ {\cal P}_{M}\psi_{-}(x){\cal P}_{M}^{\dagger}= -i\gamma^{0}\psi_{-}(t,-\vec{x}),
\end{eqnarray}
as in \eqref{CP of Majorana-1}.
The non-trivial part of 
 this deformation is the CP symmetry and, in this sense, this deformation is essentially equivalent to the formulation of the Majorana fermion with $PC={\cal P}_{M}{\cal C}_{M}$ described above. 
It is assuring that the ``parity'' defined in \eqref{trivial C and CP2} corresponds to the $\pm i\gamma^{0}$-parity and thus consistent with the classical Majorana condition. The classical Majorana condition $\psi_{+}(x)=-C\overline{\psi_{+}(x)}^{T}$ or $\psi_{-}(x)=C\overline{\psi_{-}(x)}^{T}$ in \eqref{classical Majorana}, that determines if a given fermionic field is a Majorana field, carries the same physical information as the trivial operation ${\cal C}_{M}\psi_{\pm}(x){\cal C}_{M}^{\dagger}=\psi_{\pm}(x)$ applied to the field $\psi_{\pm}(x)$ which is assumed to be the Majorana fermion $\psi_{+}(x)=-C\overline{\psi_{+}(x)}^{T}$ or $\psi_{-}(x)=C\overline{\psi_{-}(x)}^{T}$, respectively. Physics-wise, those modified C and P  symmetries do not add new ingedients to the analysis of neutron-antineutron oscillations.

\section{Discussion and conclusion}

We have shown that the use of the conventional $\gamma^{0}$-parity for the starting neutron field
gives rise to a consistent description of the emergent Majorana fermions in the oscillation process and thus a consistent description of neutron-antineutron oscillations. Physically, this choice is warranted by the fact that the neutron produced in strong interactions is viewed as a Dirac particle, with the oscillation-inducing Majorana mass terms being the effective expression of some (so far hypothetical) additional interaction. The crucial observation is that the emergent Majorana fermions are characterized by CP symmetry and are consistent with the classical Majorana condition as in \eqref{CP of Majorana-1} and \eqref{consistency condition}. Technically, the C and P defined for the starting neutron are not  good symmetries of the emergent Majorana fermions
in the present chiral description \eqref{exact-solution0} for $M_{1}\neq M_{2}$ which is required for neutron-antineutron oscillations, and thus the choice of the $\gamma^{0}$-parity
or $i\gamma^{0}$-parity for the initial neutron does not make a decisive difference (see also the Appendix \ref{A}).

Although the general Lagrangian \eqref{Lagrangian1} is physically P and CP violating, the definitions of $\gamma^0$- or $i\gamma^0$-parities still have their respective merits in the analysis of neutron oscillations: the $\gamma^0$-parity produces a criterion, in the form of the parity doubling theorem, for the existence of oscillations for particular choices of the mass parameters in \eqref{Lagrangian1} (see Ref. \cite{FT1}), while the $i\gamma^0$-parity emphasizes the P and CP invariance of the oscillation probability (see Refs. \cite{FT1,BV}).

The present formulation supports the past analyses of neutron-antineutron oscillations using the ordinary $\gamma^{0}$-parity, for example, in \cite{ft} and \cite{FT1} from a different theoretical perspective.
\\
\\

We thank M. Chaichian for helpful comments.
The present work is supported in part by JSPS KAKENHI (Grant No.18K03633).

\appendix

\section{Pauli--G\"ursey transformation}\label{A}

We show that the different choice of the parity operation, $i\gamma^{0}$ or $\gamma^{0}$,  is compensated for by the Pauli--G\"ursey transformation \cite{Pauli, Gursey, KF-PG} in the diagonalization process of the mass matrix. We thus formally understand the canonical equivalence of the two choices of the parity operation in the analysis of neutron-antineutron oscillations on the basis of an effective Lagrangian \eqref{Lagrangian1}.

In the formulation with the $i\gamma^{0}$-parity as in \cite{BV}, one may choose the Autonne--Takagi factorization   of a complex symmetric matrix (instead of \eqref{diagonal mass-1})
\begin{eqnarray}\label{Autonne--Takagi-A}
            {U^{\prime}}^{T}
            \left(\begin{array}{cc}
            m^{\dagger}_{R}& m_{D}\\
            m_{D}& m_{L}
            \end{array}\right)
            U^{\prime}
            = \left(\begin{array}{cc}
            M_{1}&0\\
            0&-M_{2}
            \end{array}\right),         
\end{eqnarray}
and  define
\begin{eqnarray} \label{variable-change-A}          
            &&\left(\begin{array}{c}
            n_{L}^{c}\\
            n_{L}
            \end{array}\right)
            = U^{\prime} \left(\begin{array}{c}
            {N}_{L}^{c}\\
            {N}_{L}
            \end{array}\right)
           ,\ \ \ \ 
            \left(\begin{array}{c}
            n_{R}\\
            n_{R}^{c}
            \end{array}\right)
            = {U^{\prime}}^{\star} 
            \left(\begin{array}{c}
            {N}_{R}\\
            {N}_{R}^{c}
            \end{array}\right).          
\end{eqnarray}
Note that the Autonne--Takagi factorization is very different from the conventional diagonalization of a hermitian matrix by a unitary transformation; the Autonne--Takagi factorization basically gives rise to characteristic values (i.e., real and positive $M_{1}$ and $M_{2}$) but the phase freedom of the diagonal elements is still left free. The total hermitian Lagrangian \eqref{Lagrangian-2} is then written as  
\begin{eqnarray}\label{exact-solution-A}
{\cal L}&=&\frac{1}{2}\{\overline{N_{L}}(x)i\delslash N_{L}(x)+\overline{N^{c}_{L}}(x)i\delslash N^{c}_{L}(x)+\overline{N_{R}}(x)i\delslash N_{R}(x)
+\overline{N^{c}_{R}}(x)i\delslash N^{c}_{R}(x)\}\nonumber\\
&-&(1/2)\left(\begin{array}{cc}
            \overline{N_{R}}&\overline{N_{R}^{c}}
            \end{array}\right)
\left(\begin{array}{cc}
            M_{1}&0\\
            0&-M_{2} 
            \end{array}\right)            
            \left(\begin{array}{c}
            N_{L}^{c}\\
            N_{L}
            \end{array}\right) +h.c.,       \nonumber\\
&=&\overline{N_{L}}(x)i\delslash N_{L}(x)+\overline{N_{R}}(x)i\delslash N_{R}(x)\nonumber\\
&-&(1/2)\{\overline{N_{R}}CM_{1}\overline{N_{R}}^{T}+N_{R}^{T}CM_{1}N_{R} -\overline{N_{L}}CM_{2}\overline{N_{L}}^{T}-N_{L}^{T}CM_{2}N_{L}
\} \nonumber\\
&=&\frac{1}{2} \overline{\psi_{+}(x)}(i\gamma^{\mu}\partial_{\mu} - M_{1})\psi_{+}(x) + 
\frac{1}{2}\overline{\psi_{-}(x)}(i\gamma^{\mu}\partial_{\mu} - M_{2})\psi_{-}(x),
\end{eqnarray}
where we defined the Majorana fields by 
\begin{eqnarray}\label{Majorana-A}
\psi_{+}(x)&=& N_{R}(x) + C\overline{N_{R}(x)}^{T}, \nonumber\\
\psi_{-}(x)&=& N_{L}(x) - C\overline{N_{L}(x)}^{T} 
\end{eqnarray}
which satisfy the classical Majorana conditions
\begin{eqnarray}\label{classical Majorana-A}
\psi_{+}(x)= C\overline{\psi_{+}(x)}^{T}, \ \ \ \psi_{-}(x)= - C\overline{\psi_{-}(x)}^{T}
\end{eqnarray}
identically in the sense that these conditions are satisfied regardless of the choice of $N_{R}(x)$ or $N_{L}(x)$.
The Lagrangian \eqref{exact-solution-A} is invariant under the CP symmetry 
\begin{eqnarray}
\ N_{L}(x) \rightarrow i\gamma^{0}C\overline{N_{L}(t,-\vec{x})}^{T},\ \ N_{R}(x) \rightarrow i\gamma^{0}C\overline{N_{R}(t,-\vec{x})}^{T}
\end{eqnarray}
 defined by $i\gamma^{0}$ parity for any real $M_{1}$ and $M_{2}$, and the same CP gives
\begin{eqnarray}
({\cal P}{\cal C})\psi_{+}(x)({\cal P}{\cal C})^{\dagger} = i\gamma^{0} \psi_{+}(t,-\vec{x}), \ \ \
({\cal P}{\cal C})\psi_{-}(x)({\cal P}{\cal C})^{\dagger} = -i\gamma^{0} \psi_{-}(t,-\vec{x})
\end{eqnarray}  
which are consistent with the classical Majorana conditions \eqref{classical Majorana-A}. According to the $i\gamma_0$-modification of the transformations \eqref{conventional C and P} which treat $N_{L}$ and $N_{R}$ as the chiral components of a Dirac field, we note, however, that neither C
\begin{eqnarray}
{\cal C}\psi_{+}(x){\cal C}^{\dagger} &=& N_{L}(x) + C\overline{N_{L}(x)}^{T}, \nonumber\\
{\cal C}\psi_{-}(x){\cal C}^{\dagger} &=&-N_{R}(x) + C\overline{N_{R}(x)}^{T}
\end{eqnarray}  
nor P ($i\gamma^{0}$-parity) 
\begin{eqnarray}
{\cal P}\psi_{+}(x){\cal P}^{\dagger} &=& i\gamma^{0} [N_{L}(x) + C\overline{N_{L}(x)}^{T}], \nonumber\\
{\cal P}\psi_{-}(x){\cal P}^{\dagger} &=& i\gamma^{0}[ N_{R}(x) - C\overline{N_{R}(x)}^{T}]
\end{eqnarray} 
are good symmetries of \eqref{exact-solution-A} for $M_{1}\neq M_{2}$, (i.e., $|m|\neq 0$ in \eqref{Mass eigenvlaues4} using the fact to be mentioned below). Nevertheless the Lagrangian \eqref{exact-solution-A} is physically P and C invariant, upon a redefinition of these transformations which complies with the Majorana condition \eqref{Majorana-A}, using the deformed symmetry generators as in \eqref{Deformed C and P-2}.
 
When one remembers that the starting mass matrix and the neutron field are common and 
the mass eigenvalues $M_{1}$ and $M_{2}$ are common for either choice of the parity operation, as is directly confirmed by deriving the relations  \eqref{Mass eigenvlaues1} and \eqref{Mass eigenvlaues2} for the relation \eqref{Autonne--Takagi-A} also with $U\rightarrow U^{\prime}$, one can confirm that the matrix $U^{\prime}$ in \eqref{Autonne--Takagi-A} is written using $U$ in \eqref{diagonal mass-1} as
 \begin{eqnarray}
 U^{\prime}=UU_{0}
 \end{eqnarray}
 with 
 \begin{eqnarray}
 U_{0}=e^{-i\pi/4}\left(\begin{array}{cc}
            1&0\\
            0&i
            \end{array}\right) =\left(\begin{array}{cc}
            e^{-i\pi/4}&0\\
            0&e^{i\pi/4}
            \end{array}\right).
 \end{eqnarray}
The mass eigenstates in \eqref{variable-change-A}  and the mass eigenstates in \eqref{variable-change-1} are then related by
\begin{eqnarray}
U_{0}\left(\begin{array}{c}
            {N}_{L}^{c}\\
            {N}_{L}
            \end{array}\right)_{i\gamma^{0}}
            =\left(\begin{array}{c}
            {N}_{L}^{c}\\
            {N}_{L}
            \end{array}\right)_{\gamma^{0}}
             ,\ \ \
            {U_{0}}^{\star} 
            \left(\begin{array}{c}
            {N}_{R}\\
            {N}_{R}^{c}
            \end{array}\right)_{i\gamma^{0}}
            =\left(\begin{array}{c}
            {N}_{R}\\
            {N}_{R}^{c}
            \end{array}\right)_{\gamma^{0}}         
\end{eqnarray}
with $U_{0}\in U(2)$ of the Pauli--G\"ursey canonical transformation \cite{Pauli, Gursey, KF-PG}.  In this sense, the two different definitions of parity are canonically equivalent. The secret of the appearance of this relation is traced to the hidden freedom in the definition of classical Majorana fermions; the definition of  Majorana fermions in \eqref{Majorana-A} is extended to the definition of Majorana fermions in \eqref{Majorana} with a phase freedom, which is precisely related to this freedom of the canonical transformation. To be precise, it is confirmed that
\begin{eqnarray}
\{\psi_{\pm}(x)\}_{i\gamma^{0}} =  (-i)\{\psi_{\pm}(x)\}_{\gamma^{0}}
\end{eqnarray}
in the present phase convention, which is consistent with  the classical Majorana conditions \eqref{classical Majorana} and \eqref{classical Majorana-A}. It is confirmed that the same oscillation formula as in \eqref{oscillation-1} is valid for the description with $i\gamma^{0}$ parity by noting $|(\hat{U}_{21}\hat{U}^{\star}_{11})|^{2}=|(U_{21}U^{\star}_{11})|^{2}=|(U^{\prime}_{21}{U^{\prime}}^{\star}_{11})|^{2}$. The CP invariance corresponds to $U^{\prime}={U^{\prime}}^{\star}$. 

Finally, we would like to add a comment on the formulation in \cite{BV}  from our present point of view.
We recall the use of a specific Pauli-Gursey transformation  in the context of a seesaw model of neutrinos \cite{Fujikawa}, which is relevant to the present problem. As for a related use of the Bogoliubov transformation, see \cite{FT-1,FT-neutrino}.
 One may apply the Pauli--G\"ursey $U(2)$ transformation 
\begin{eqnarray}
\left(\begin{array}{c}
            {N}_{L}^{c}\\
            {N}_{L}
            \end{array}\right)
            =O\left(\begin{array}{c}
            {\tilde{n}}_{L}^{c}\\
            {\tilde{n}}_{L}
            \end{array}\right)
             ,\ \ \
            \left(\begin{array}{c}
            {N}_{R}\\
            {N}_{R}^{c}
            \end{array}\right)
            =O\left(\begin{array}{c}
            {\tilde{n}}_{R}\\
            {\tilde{n}}_{R}^{c}
            \end{array}\right)        
\end{eqnarray}
with an element 
\begin{eqnarray}
O=\frac{1}{\sqrt{2}}\left(\begin{array}{cc}
            1&1\\
            -1&1
            \end{array}\right)
\end{eqnarray}
to the first line of the Lagrangian \eqref{exact-solution-A}, which corresponds to a single generation model of the neutrino, one obtains
\begin{eqnarray}\label{exact-solution-A-2}
{\cal L}
&=&1/2\{\overline{\tilde{n}}(x)i\delslash \tilde{n}(x)+\overline{\tilde{n}^{c}}(x)i\delslash \tilde{n}^{c}(x)\}\nonumber\\
&-& 1/4\{\overline{\tilde{n}}(M_{1}+M_{2})\tilde{n}+\overline{\tilde{n}^{c}}(M_{1}+M_{2})\tilde{n}^{c}\}\nonumber\\ &-&1/4\{\overline{\tilde{n}}(M_{1}-M_{2})\tilde{n}^{c} +\overline{\tilde{n}^{c}}(M_{1}-M_{2})\tilde{n},
\} 
\end{eqnarray}
which is invariant under C, P($i\gamma^{0}$ parity) and CP. See eq. (45) and eqs. (55)-(57) in \cite{Fujikawa}. If one sets $M_{1}+M_{2}=2M$ and $M_{1}-M_{2}=2\epsilon$,
this Lagrangian becomes
\begin{eqnarray}\label{exact-solution-A-3}
{\cal L}
&=&\overline{\tilde{n}}(x)i\delslash \tilde{n}(x) -
M\overline{\tilde{n}}\tilde{n} -\frac{1}{2}\epsilon[\overline{\tilde{n}}C
\overline{\tilde{n}}^{T} +\tilde{{n}}^{T}C\tilde{n}],
\end{eqnarray}
which has precisely the form of the Lagrangian discussed in \cite{BV}.  

This shows that our starting Lagrangian \eqref{Lagrangian1} and our Lagrangians \eqref{exact-solution0} and \eqref{exact-solution-A}, and the Lagrangian used in \cite{BV} are all related by the Pauli--G\"ursey transformation and thus canonically equivalent. This fact, however, does not imply that our starting Lagrangian \eqref{Lagrangian1} is C, P ($i\gamma^{0}$ parity) and CP invariant since the Pauli--G\"ursey unitary transformations can carry these symmetries away to the interaction sector depending on the model. Let us note that, in the context of neutron oscillations, the Lagrangian \eqref{exact-solution-A-3} does not make the analysis of discrete transformations more transparent than the diagonal Majorana Lagrangian \eqref{exact-solution-A}.  
Moreover, the parameter $M$ in \eqref{exact-solution-A-3} is not the neutron mass $m_D$, but a complicated relation between the mass parameters in \eqref{Lagrangian1} (calculated exactly in \cite{FT1}), making the meaning of $M$ phenomenologically less transparent. There is absolutely no reason why one should adopt the Lagrangian \eqref{exact-solution-A-3}, which is one of the canonically equivalent Lagrangians, to describe the neutron-antineutron oscillations.

Incidentally, one can confirm that the $\gamma^{0}$-parity becomes good symmetry for $M_{1}=M_{2}$ in either \eqref{exact-solution0} or \eqref{exact-solution-A} (and also in \eqref{exact-solution-A-2}) and thus leading to the parity-doublet theorem, namely, no oscillations \cite{FT1}.

\section{Direct parity analysis of $\Delta B=2$ Lagrangian}\label{B}

The Lagrangian \eqref{Lagrangian1} was systematically analyzed in \cite{FT1} regarding the physical P, C and CP symmetries, before and after the diagonalization. Here we give a summary of those results. One can easily bring the Lagrangian \eqref{Lagrangian1} to the form
\begin{eqnarray}\label{3}
{\cal L}&=&\overline{n}(x)i\gamma^{\mu}\partial_{\mu}n(x) - m_D\overline{n}(x)n(x)\nonumber\\
&-&\frac{i}{2}|m|[e^{i\alpha}n^{T}(x)Cn(x) - e^{-i\alpha}\overline{n}(x)C\overline{n}^{T}(x)]\nonumber\\
&-&\frac{i}{2}|m_5|[n^{T}(x)C\gamma_{5}n(x) + \overline{n}(x)C\gamma_{5}\overline{n}^{T}(x)],
\end{eqnarray}
where $m_D$ and $\alpha$ are real parameters, by a redefinition of the neutron field which absorbs the phase of $m_5$. No other redefinitions of the neutron field are permitted due to the baryon number symmetry violation.

One can check directly that, under $\gamma_0$-parity definition of the neutron field $n$, the $|m|$-term in \eqref{3} is parity odd, while the $|m_5|$-term is even; under $i\gamma_0$-parity, the $|m|$-term in \eqref{3} is parity even, while the $|m_5|$-term is odd. Actually, any definition of the parity with an arbitrary phase, $e^{i\varphi}\gamma_0$-parity, leads to a result of parity violation. The charge conjugation is generally not conserved either. One confirms that, irrespective of the value of $\alpha\neq 0$ in \eqref{3}, the CP violation cannot be eliminated as long as $m,m_5\neq0$. Consequently, for $\alpha\neq0$ and $m m_5\neq 0$ in \eqref{3}, parity and CP are intrinsically violated.  

Upon diagonalization, the Lagrangian \eqref{3} is brought to the form of a sum of two free Majorana Lagrangians (see eq.(40) in \cite{FT1}), 
$$
{\cal L}=\frac{1}{2} \overline{\psi_{+}(x)}(i\gamma^{\mu}\partial_{\mu} - M_{1})\psi_{+}(x) + 
\frac{1}{2}\overline{\psi_{-}(x)}(i\gamma^{\mu}\partial_{\mu} - M_{2})\psi_{-}(x),
$$
where
\begin{eqnarray}\label{21}
M_{1,2}&=&\left(\left[{\sqrt{m_{D}^{2} + |m_{5}|^{2}}} \pm |m|\sqrt{1-(\tilde {|m|}/|m|)^{2}}\right]^{2}+(\tilde{|m|})^{2}\right)^{1/2},
\end{eqnarray}
with $\tilde {|m|}\equiv|m|\sin\alpha\sin2\phi$ and $\sin2\phi\equiv |m_{5}|/\sqrt{m_D^{2}+|m_{5}|^{2}}$.

On the other hand, the exact solution of the relation \eqref{Mass eigenvlaues2} is written for the parameterization of mass parameters in \eqref{3} as
\begin{eqnarray}\label{exact mass formula}
M^{2}_{1,2}&=&m_{D}^{2}+|m_{5}|^{2}+|m|^{2}\pm 2|m|\sqrt{m_{D}^{2}+|m_{5}|^{2}\cos^{2}\alpha}.
\end{eqnarray}
One can confirm that \eqref{21} agrees exactly with \eqref{exact mass formula}. This shows that the characteristic values of the mass matrix are invariant under the canonical transformation, and the oscillation formula in \cite{FT1} is exact and agrees with \eqref{oscillation-1}, as it should be.

In terms of the original parameters in the Lagrangian \eqref{Lagrangian1}, we note that the mass splitting between the Majorana fields, and consequently the oscillation, can occur only if $m\neq 0$.  
Taking into account that in the physically relevant situation $|m|,|m_5|\ll m_D$, one finds from \eqref{21} in this approximation
$|M_{1} - M_{2}|=2|m|$, %(see also formula \eqref{Mass eigenvlaues4} above)
namely the correction to the value of the probability of oscillation due to the CP-breaking parameters $\alpha$ and $m_5$, though not exactly vanishing, is immaterial. However, for any value of the parameters in the Lagrangian \eqref{Lagrangian1}, the probability of oscillation {\it per se} is P and CP invariant. In other words, although the general Lagrangian is P and CP violating, the oscillation probability is not.

The choice of the conventional $\gamma^{0}$-parity transformation is as legitimate as the choice of the $i\gamma^{0}$-parity transformation advocated as the only sensible one in Ref. \cite{BV}, since the general $\Delta B=2$ Lagrangian \eqref{Lagrangian1} written in terms of the original neutron field $n(x)$ is {\it not invariant} under any of them as long as $m m_5\neq 0$, and in general is not invariant under any conceivable $e^{i\varphi}\gamma_0$-parity transformation  (see \cite{FT1}). As a result, the Lagrangian \eqref{Lagrangian1} describes effectively mass-generating interactions which break parity, and in general also CP (unless $\alpha$ is set to zero in \eqref{3} by hand, or by physical arguments). This breaking is not observable in neutron-antineutron oscillations, which is based on the simple quadratic Majorana Lagrangian \eqref{exact-solution0}, but it is carried over to the interaction terms by the mixing matrix $U$. Moreover, as shown in \cite{FT1}, if high-energy CP-violating interactions generate the $m_5$-term in the Lagrangian \eqref{Lagrangian1}, their effects would show up in an enhanced electric dipole moment of the neutron compared to the one generated by the QCD $\theta$-term. In short, no choice of the parity transformation of the original neutron field $n(x)$ gives more information than any other, when the total Lagrangian, with all the interactions included, is taken into consideration.

\end{document}